\documentclass[5p,times]{elsarticle}
\usepackage{lipsum}
\usepackage{lineno,hyperref}
\modulolinenumbers[5]
\usepackage[pdftex]{color}
\usepackage[font=footnotesize,labelfont=bf]{caption}
\usepackage[font=footnotesize,labelfont=bf]{subcaption}
\journal{Journal of \LaTeX\ Templates}

\usepackage{colortbl}
\usepackage{xcolor}

\begin{document}

\title{Antagonistic coinfection in rock-paper-scissors models during concurrent epidemics}

\author[zuyd]{J. Menezes\corref{mycorrespondingauthor}}
\cortext[mycorrespondingauthor]{Corresponding author}
\ead{josinaldo.menezes@zuyd.nl}
\address[zuyd]{Research Centre for Data Intelligence, Zuyd University of Applied Sciences, Paul Henri Spaaklaan 3D, 6229 EN, Maastricht, The Netherlands}

\author[ufpb]{R. Menezes}
\ead{rmenezes@dcx.ufpb.br}
\address[ufpb]{Departament of Exact Sciences, Federal University of Paraiba, 58297-000 Rio Tinto, PB, Brazil}

\author[ect-ufrn]{S. Batista}
\ead{simone.batista@ufrn.br}
\address[ect-ufrn]{School of Science and Technology, Federal University of Rio Grande do Norte,
59072-970, P.O. Box 1524, Natal, RN, Brazil}

\author[deb-ufrn]{E. Rangel} 
\ead{enzo.rangel.107@ufr.edu.br}
\address[deb-ufrn]{Department of Computer Engineering and Automation, Federal University of Rio Grande do Norte, Av. Senador Salgado Filho 300, Natal, 59078-970, Brazil}

\begin{abstract}
We investigate the dynamics of dual disease epidemics within the spatial rock-paper-scissors model. In this framework, individuals from all species are equally susceptible to infection by two distinct pathogens transmitted via person-to-person contact. We assume antagonistic mortality, where the simultaneous occurrence of coinfection reduces the probability of host mortality due to complications arising from either coexisting disease. Specifically, we explore two scenarios: global antagonism, where the presence of one pathogen inhibits the progression of the other in coinfected hosts, and uneven antagonism, where only one pathogen affects the development of the other. 
Using stochastic simulations, we show that the characteristic length scale of the spatial patterns emerging from random initial conditions diminishes as antagonism becomes more significant.
 We find that antagonism enhances species population growth and reduces the average probability of healthy organisms becoming infected.
Additionally, introducing individuals' mobility restrictions significantly decreases both organisms' infection risk and selection pressures.
Our results demonstrate that combining mobility restrictions with antagonistic coinfection can increase organisms' life expectancy by up to $54\%$. Our findings show that integrating antagonistic coinfection and mobility restriction strategies into ecological models may provide insights into designing interventions for managing concurrent epidemics in complex systems.
\end{abstract}

\maketitle

\section{Introduction}
\label{sec1}

It is well established that space plays a crucial role in species coexistence. In biological systems characterised by local cyclic competition for resources, stability arises due to the emergence of distinct spatial domains \cite{Coli,Allelopathy}. For instance, this phenomenon has been observed in experiments involving strains of \textit{Escherichia coli}, where the classic rock-paper-scissors model captures cyclic dominance \cite{bacteria}. Similarly, local cyclic interactions have been shown to ensure coexistence in other ecosystems \cite{lizards,coral}. Consequently, this model has become a cornerstone for simulating diverse biological scenarios and testing hypotheses \cite{mobilia2,mobilia3,uneven,tanimoto2,park2023,rps-ambush,rps-scripta}.

Beyond ecological dynamics, research into behavioural strategies to mitigate viral disease spread has gained significant attention \cite{social1,disease4,disease3,disease2}. For example, global implementation of social distancing measures has demonstrated individual and collective benefits \cite{socialdist,soc,plasticity2}. Additionally, mobility restrictions have proven effective in reducing infection rates and minimising the impact of diseases on populations \cite{mr0,mr1,mr2,10.1371/journal.pone.0254403}. However, as pathogens evolve through mutation, adopting such strategies is imperative to address changing transmission dynamics and mortality rates \cite{CAPAROGLU2021111246}.
This adaptability highlights the importance of flexible spatial organisation in controlling organism dispersal \cite{plasticity2,Gene,mutating1,plasticity1}. 

A recent study investigated the concurrency of two epidemics in spatial cyclic models, assuming a synergistic increase in the risk of coinfected hosts dying due to complications from either disease \cite{synergy}. The formation of single-disease spatial domains bordered by interfaces of coinfected hosts, driven by curvature dynamics, was observed. Furthermore, it was reported that organisms' infection risk could be maximised or minimised depending on the level of synergy acting in coinfected individuals, directly influencing species population dynamics.

Despite the scientific effort to understand the relationship between pathogens and hosts, the study of antagonism in multiple epidemics within cyclic games has not been studied \cite{coifectedhiv}.
Specifically, one infection inhibits the development of the other, acting as a protective mechanism that shields the host from the detrimental effects of the secondary virus. This interaction results in a decreased effective mortality rate \cite{antag2}. Antagonism, the opposite of synergy in epidemiological terms \cite{antag3}, has been studied at both individual and population levels, with literature highlighting its effects on host health and spatial population distribution, as well as its broader implications for ecosystem dynamics \cite{antag4,antag1}.

Given the significance of antagonism in ecosystem dynamics, this study explores the interplay between antagonistic coinfection and mobility restriction strategies within a stochastic rock-paper-scissors model featuring two concurrent epidemics \cite{epidemicbook,tanimoto}. These pathogens interact antagonistically, reducing the mortality risk for coinfected organisms developing diseases caused by both viruses.
We investigate population-level dynamics by observing and quantifying changes in spatial patterns and computing the impact on species densities. 

Simulations are conducted across scenarios where antagonism affects coinfected individuals differently: globally, by lowering the mortality rate for both diseases or unevenly, where only the mortality risk associated with one disease is reduced while the other remains constant.
Using stochastic simulations, we analyse the effects of antagonistic coinfection on individual-level outcomes, quantifying organisms' cure probability and infection risk. Furthermore, we estimate the average survival time of individuals, accounting for the combined effects of antagonism and mobility restrictions on selection risk.

The outline of this paper is as follows: in Sec.~\ref{sec2}, we introduce our stochastic model and describe the simulations. The spatial patterns are investigated in 
Sec.~\ref{sec3}, and characteristic length scale in computed in 
Sec.~\ref{sec4}. We study the effects of the antagonistic infection in the organisms' infection risk and species densities in Sec.~\ref{sec5}. We then investigate the benefits of the mobility restriction strategy in reducing the infection and selections and increasing the organisms' expected survival time in Sec.~\ref{sec6}.
Finally, Sec. \ref{sec7} highlights our conclusions and discussions.

\section{The model}
\label{sec2}

\begin{figure}
\centering
\includegraphics[width=40mm]{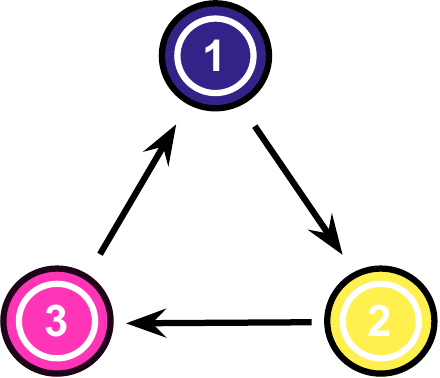}
\caption{Illustration of the rock-paper-scissors model. Selection interactions are represented by arrows denoting the dominance of organisms of species $i$ over individuals of species $i+1$.}
\label{fig1}
\end{figure}

We explore the stochastic model of rock-paper-scissors dynamics, where three species compete according to the rules: scissors cut paper, paper wraps rock, and rock crushes scissors. This is illustrated in Fig.~\ref{fig1}, where each species is denoted by i, ranging from $1$ to $3$, and is identified by $i=i+3\alpha$, where $\alpha$ is an integer. We conduct simulations of two concurrent epidemics, wherein the pathogens responsible for the diseases can spread from person to person.
All organisms within each species are equally susceptible to independent contamination by both pathogens, allowing for coinfection at any stage. Upon infection, an individual becomes ill, potentially leading to death. Moreover, recovering from illness does not confer immunity against reinfection.
In cases of coinfection, the decline in the host's health may be slowed because the presence of one pathogen can interfere with the multiplication of the other, thereby mitigating the overall detrimental effects on the host's well-being. The antagonistic interaction during coinfection decreases the mortality rates of the diseases, thus prolonging the hosts' lives and giving them more opportunities to recover.

\subsection{Simulations}

We run stochastic simulations to explore the scenarios of propagating two distinct disease epidemics within a cyclic spatial game. Our approach follows the May-Leonard framework, which disregards a conservation law for the total number of individuals \cite{leonard}. This means that the total number of individuals is not conserved.
The simulations are carried out on square lattices with periodic boundary conditions, where the maximum number of organisms is denoted by $\mathcal{N}$, corresponding to the total number of grid points, with at most one individual occupying each grid site.
We define the density of individuals of species $i$, $\rho_i(t)$, with $i=1,2,3$, as the proportion of the lattice occupied by individuals of species $i$ at time $t$: \begin{equation} \rho_i(t)=\frac{I_i(t)}{\mathcal{N}}, \end{equation} where $I_i(t)$ represents the total number of organisms of species $i$ at time $t$.

We establish random initial conditions by placing each organism at a randomly selected grid point. We assume an equal initial density for every species, approximately $\rho_i \approx \mathcal{N}/3$, for $i=1,2,3$, and the initial number of individuals is roughly the maximum integer value that can occupy the lattice, $I_i (t=0) \approx \mathcal{N}/3$, with $i=1,2,	3$ - while the remaining grid sites are vacant in the initial state. Additionally, for each species, the initial proportion of individuals infected by each pathogen is assumed to be 3\%.

We use several notations to characterise the states of individuals within a population dynamics model. Let $h_i$ denote healthy individuals of species $i$, where $i$ can take values from $1$ to $3$. Sick organisms carrying pathogens $1$ or $2$ are represented as $s^{1}_i$ and $s^{2}_i$, respectively, while $s^{1,2}_i$ represents individuals coinfected by both pathogens. The general notation $i$ refers to all individuals, regardless of their health status.

We outline the stochastic interactions as follows:

\begin{itemize}
\item 
Selection: $$ i\ j \to i\ \otimes\,,$$ with $ j = i+1$ and $\otimes$ representing an empty space. This interaction results in the removal of the individual of species $i+1$ from the grid point, leaving the sites vacant.
\item
Reproduction: $$ i\ \otimes \to i\ i\,.$$ An individual of species i reproduces, filling an adjacent empty site with another individual of the same species.
\item 
Mobility: $$ i\ \odot \to \odot\ i\,,	$$ where $\odot$ represents either an individual of any species or an empty site. During mobility, an organism swaps position with another individual or an empty space.
\item 
Infection by disease $1$: $$s^{1}_i\ h_j \to s^{1}_i\ s^{1}_j\,$$ $$s^{1,2}_i\ h_j \to s^{1,2}_i\ s^{1}_j\,$$
$$s^{1}_i\ s^{2}_j \to s^{1}_i\ s^{1,2}_j\,$$ $$s^{1,2}_i\ s^{2}_j \to s^{1,2}_i\ s^{1,2}_j\,$$ with $i,j=1,2,3$. An organism of species $i$ carrying the pathogen $1$ contaminates a healthy individual or a host of pathogen $2$.
\item 
Infection by disease $2$: $$s^{2}_i\ h_j \to s^{2}_i\ s^{2}_j\,$$ $$s^{1,2}_i\ h_j \to s^{1,2}_i\ s^{2}_j\,$$
$$s^{2}_i\ s^{1}_j \to s^{2}_i\ s^{1,2}_j\,$$ $$s^{1,2}_i\ s^{1}_j \to s^{1,2}_i\ s^{1,2}_j\,$$ with $i,j=1,2,3$. An ill individual of species $i$ transmits the pathogen $2$ to a healthy individual or a host of pathogen $1$.
\item 
Cure of disease $1$: $$s^{1}_i \to h_i\,$$ $$s^{1,2}_i \to s^{2}_i\,$$. An individual recovers from infection by pathogen $1$, becoming healthy or retaining only pathogen $2$.

\item 
Cure of disease $2$: $$s^{2}_i \to h_i\,$$ $$s^{1,2}_i \to s^{1}_i\,$$ When cured from disease $2$, the organism becomes healthy or
still sick due to pathogen $1$.
\item 
Death caused by disease $1$: $$s^{1}_i \to \otimes\,$$ $$s^{1,2}_i \to \otimes\,$$ Ill individuals succumb to disease $1$, leading to vacant grid sites.
\item
Death due to disease $2$: $$s^{2}_i \to \otimes\,$$ $$s^{1,2}_i \to \otimes\,$$ Similarly, disease $2$ can cause the death of infected individuals, resulting in empty grid points.
\end{itemize}

The randomness of interactions is governed by a set of probabilities, which are computed using key parameters: $S$ for selection, $R$ for reproduction, $M$ for mobility, $T$ for pathogen transmissibility and $C$ for cure. Regarding the 
disease mortality, the death rate in sick organisms infected by a single pathogen is controlled by the parameter $\mu$. However, due to the antagonistic action of one pathogen on the effects of the other, the chances of an individual dying from the harmful effects of one of the diseases decrease when coinfected.

To account for this, we introduce antagonistic factors, $\gamma_1$ and $\gamma_2$, for diseases $1$ and $2$, respectively. These are real parameters ranging from $0$ to $1$, which measures the intensity of antagonism in coinfected individuals. 
This means that the mortality rates of diseases $1$ and $2$ are rescaled by the factors $1-\gamma_1$ and $1-\gamma_2$, respectively. 

For instance, in the case where coinfection exhibits no antagonism for either disease ($\gamma_1=\gamma_2=0$), the mortality rates of both diseases in coinfected individuals are identical to those observed in singly infected hosts. Consider another example where the level of antagonism is $70\%$ for both diseases ($\gamma_1=\gamma_2=0.7$), 
reducing the diseases' mortality in coinfected hosts to $0.3\,\mu$. In the extreme case of total antagonistic coinfection ($\gamma_1=\gamma_2=1.0$), one pathogen fully inhibits the effects of the other, thereby abolishing the likelihood of coinfected organisms succumbing to disease-related complications.

Finally, there is also the case of uneven antagonism. For example, if the presence of pathogen $2$ inhibits the spread of pathogen $1$, while pathogen $1$ does not affect pathogen $2$.
As an example, 
$\gamma_1=0.7$ and $\gamma_2=0.0$ 
indicate that the mortality risk for a coinfected host due to disease $1$ is $30\%$ of that associated with disease $2$.
For simplification, when antagonism affects both diseases equally, we denote the antagonism factors as $\gamma_1=\gamma_2=\gamma$.

The implementation operates within the Moore neighbourhood, enabling each organism to interact with any of its eight immediate neighbours. The algorithm 
proceeds as follows: i) a random selection of an active individual from all organisms on the lattice; ii) a random selection of one interaction to execute; iii) the designation of one of the eight immediate neighbours for the action to occur.
Successful execution occurs when active individuals can carry out the intended action over passive ones, in accordance with the model's rules, thereby advancing time by one step. In cases where this isn't feasible, the algorithm iterates through these steps until a viable interaction is achieved. The completion of $\mathcal{N}$ timesteps constitutes the end of one generation – our designated unit of time.

\begin{figure*}
	\centering
	 \begin{subfigure}{.24\textwidth}
    \centering
    \includegraphics[width=42mm]{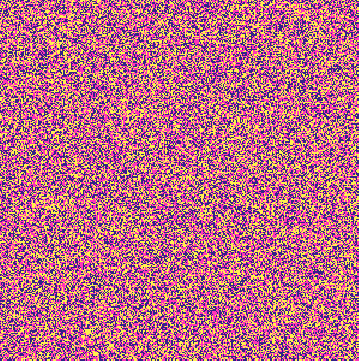}
    \caption{}\label{fig2a}
  \end{subfigure} 
  \begin{subfigure}{.24\textwidth}
    \centering
    \includegraphics[width=42mm]{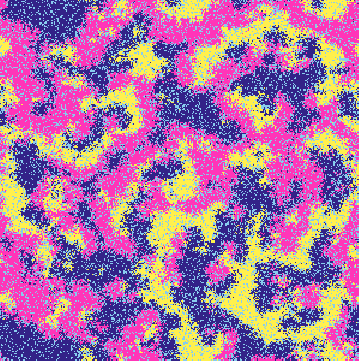}
    \caption{}\label{fig2b}
  \end{subfigure} %
  \begin{subfigure}{.24\textwidth}
    \centering
    \includegraphics[width=42mm]{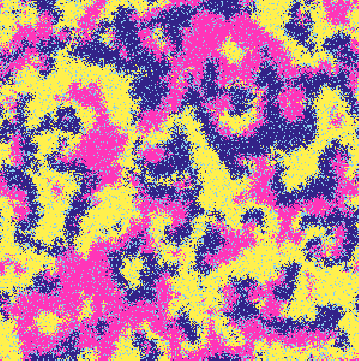}
    \caption{}\label{fig2c}
  \end{subfigure} 
           \begin{subfigure}{.24\textwidth}
    \centering
    \includegraphics[width=42mm]{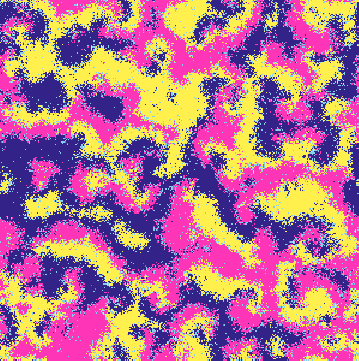}
    \caption{}\label{fig2d}
  \end{subfigure}
 \caption{Snapshots of the rock-paper-scissors model with antagonistic coinfection. These snapshots capture the spatial organisation of organisms on a lattice with $300^2$ grid sites, evolving over 3000 generations. Figure \ref{fig2a} presents the initial random conditions, while Figs.~\ref{fig2b}, \ref{fig2c}, and \ref{fig2d} showcase the spatial distribution of individuals at the end of Simulations A ($\gamma=0.0$), B ($\gamma=0.5$), and C ($\gamma=1.0$), respectively. Purple, yellow, and pink dots represent individuals of species $1$, $2$, and $3$, while light blue dots indicate empty spaces.
}
 \label{fig2}
\end{figure*}
\begin{figure}
\centering
    \includegraphics[width=85mm]{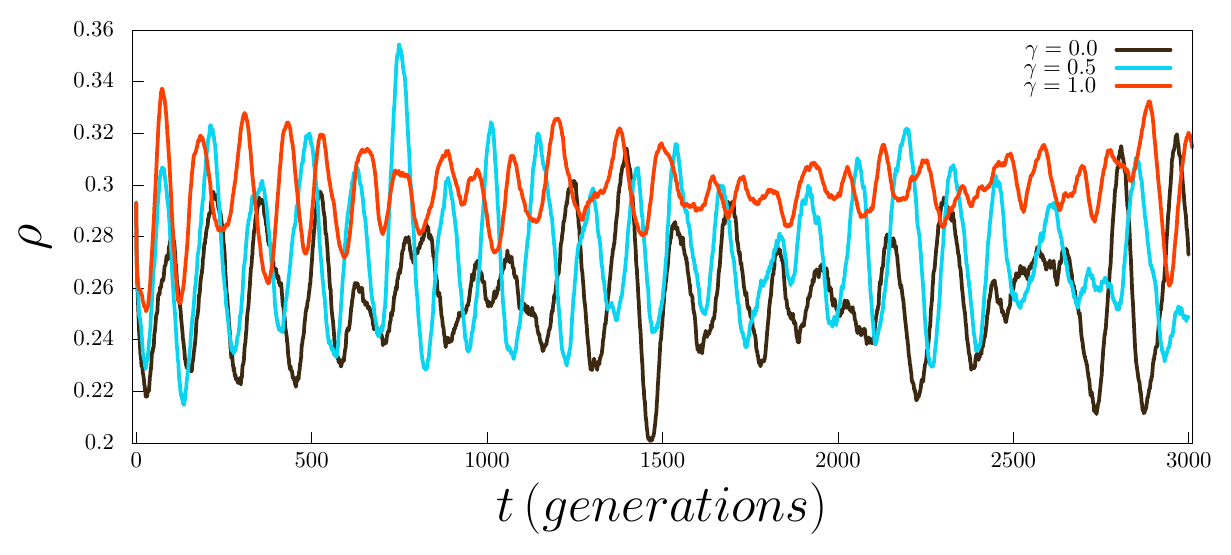}
\caption{Temporal variation of species densities in Simulations A, B, and C, depicted in the snapshots shown in Figs. \ref{fig2b} (brown line), \ref{fig2c} (cyan line), and \ref{fig2d} (red line).}
	\label{fig3}
\end{figure}
\section{Spatial patterns}
\label{sec3}

We begin by investigating the spatial arrangements resulting from random initial conditions while modulating the mortality rate of coinfected individuals. The antagonism factor is assumed to be identical for both diseases' mortality, assumed as follows:
\begin{itemize}
\item
Simulation A: $\gamma=0.0$, indicating no antagonistic effect on the death rate for coinfected individuals such that their mortality probability is identical to that of an individual infected with only one specific disease.

\item
Simulation B: $\gamma=0.5$, where a coinfected host has half the likelihood of dying from one of the diseases, compared to the scenario of a single infection.
\item
Simulation C: $\gamma=1.0$, representing a scenario where the antagonistic effect on the mortality rate of coinfected organisms is total, effectively reducing the chances of death by $100\%$.
\end{itemize}
To facilitate this exploration, we conducted a single simulation on a lattice comprising $300^2$ grid points for a timespan of $3000$ generations. The simulations parameters were set as follows: $S=R=1.0$, $M=3.0$, $T=4.0$, and $\mu=C=0.1$.

Figure \ref{fig2a} illustrates the random initial conditions employed for the simulations, while Figs. \ref{fig2b}, \ref{fig2c}, and \ref{fig2d} display snapshots of the spatial organisation of organisms at the end of Simulations A, B, and C, respectively. As indicated in Fig. \ref{fig1}, purple, yellow, and pink dots represent individuals of species $1$, $2$, and $3$, irrespective of their health status—whether healthy, infected with a single pathogen, or coinfected. Light blue dots indicate empty spaces.

As depicted in Figures \ref{fig2b} to \ref{fig2d}, individuals of the same species cluster into distinct spatial domains following an initial phase of pattern formation. The cyclic selection rules inherent in the rock-paper-scissors game lead to the emergence of spirals emerging from random initial conditions. Notably, as $\gamma$ increases, reflecting a reduced likelihood of death for coinfected individuals, the number of empty spaces (depicted as light blue dots) decreases in Fig.~\ref{fig2c} compared to Fig.~\ref{fig2b}, where no antagonistic effect is present. This trend is further highlighted in Fig.~\ref{fig2d}, where no coinfected individual succumbs to complications from either disease. Furthermore, due to antagonistic mortality, most empty spaces concentrate along the borders between spiral arms. This phenomenon arises because, apart from vacancies resulting from the death of individuals infected with a single pathogen, the most significant fraction of empty spaces within the system arises from selection interactions.

\subsection{Dynamics of species densities}

We also computed the temporal fluctuations in the total species density $\rho$, irrespective of the proportions of sick individuals from single pathogens, coinfected cases, and healthy ones. The species density is depicted in Fig.~\ref{fig3}, with brown, cyan, and red representing the densities of species $1$ for Simulations A, B, and C, respectively. The choice of species $1$ was arbitrary as the average densities for all species are identical due to the uniform application of the stochastic model rules and probabilities across species.

The results reinforce the observations made in Figs. ~\ref{fig2b} to \ref{fig2d}: for a highly
antagonistic mortality scenario ($\gamma\,\to\,1$), the density of empty spaces decreases. 
This happens because the density of organisms of each species, which remains on average constant over time, increases as the risk of death diminishes ($\gamma \to 1$) due to the reduction of the empty spaces - that become more concentrated on the borders between domains occupied by species $i$ and $i+1$, as observed in Fig. \ref{fig2}.  
Consequently, the red line (Simulation A) typically exhibits the highest density of empty spaces, while scenarios without antagonism demonstrate the lowest density.

Furthermore, the findings in Fig.~\ref{fig3} also suggest that the periodicity in organism densities is more pronounced for $\gamma=1.0$. This phenomenon occurs because the reduced number of deaths caused by coinfection enables more organisms to engage actively in spatial dynamics, thereby amplifying the changes inherent to cyclic dominance governed by the rock-paper-scissors rules. Thus, increasing the frequency of fluctuations in species density. 
\begin{figure}
\centering
    \includegraphics[width=85mm]{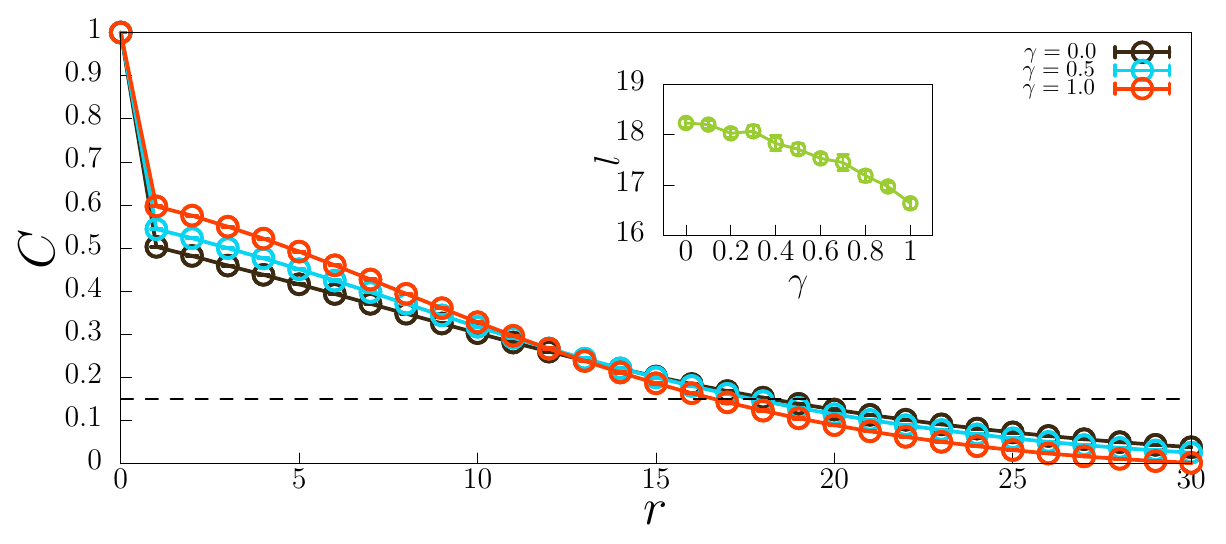}
\caption{Autocorrelation function for organisms of the same species for various coinfection antagonistic factors. The results were averaged over 100 simulations conducted on lattices with $500^2$ grid sites until $t=5000$ generations. The error bars show the standard deviation. Brown, cyan, and red lines correspond to the results for $\gamma=0.0$, $\gamma=0.5$, and $\gamma=1.0$, respectively. The inset figure illustrates the characteristic length scale as a function of the coinfection antagonistic factor. The horizontal dashed line indicates the threshold assumed to calculate the
characteristic length scale shown in the inner panel}
	\label{fig4}
\end{figure}

\section{Spatial Autocorrelation Function and  Characteristic Length Scale}
\label{sec4}

In this section, we explore the complex interplay between antagonistic coinfection and the spatial organisation of individuals within species, focusing on the characteristic scale of spatial domains. Our investigation begins with a detailed examination of the spatial autocorrelation function, denoted as $C_i(r)$, in relation to the radial coordinate $r$ for individuals belonging to each species
($i=1,2,3$) \cite{plasticity,mobilityrestrictions}.

To characterise the spatial distribution of species $i$ within the lattice, we introduce the function $\phi_i(\vec{r})$. By utilising the mean value $\langle\phi_i\rangle$ for the Fourier transform:
\begin{equation}
\varphi_i(\vec{K}) = \mathcal{F}\{\phi_i(\vec{r})-\langle\phi_i\rangle\}, 
\end{equation}
and defining the spectral densities as 
\begin{equation}
S_i(\vec{K}) = \sum_{K_x, K_y}\varphi_i(\vec{K}),
\end{equation}

We subsequently perform a normalised inverse Fourier transform to extract the autocorrelation function for species $i$:
\begin{equation}
C_i(\vec{r}') = \frac{\mathcal{F}^{-1}\{S_i(\vec{K})\}}{C(0)},
\end{equation}
which we express as a function of $r$:
\begin{equation}
C_i(r') = \sum_{|\vec{r}'|=x+y} \frac{C_i(\vec{r}')}{\min\left[2N-(x+y+1), (x+y+1)\right]}.
\end{equation}
To determine the characteristic length scale $l_i$ for the spatial domains of species $i$, we define the threshold $C_i(l_i)=0.15$. Given the uniformity of the autocorrelation function across species, we use data from species $1$ to infer the characteristic length scale for organisms of all species, denoting it as $C$ and $l$.

Figure \ref{fig4} displays the autocorrelation function $C$ for varying $\gamma$, depicted as $\gamma=0.0$ (brown line), $\gamma=0.5$ (cyan line), and $\gamma=1.0$ (red line). These results were obtained from from an ensemble of $100$ simulations conducted on lattices consisting of $500^2$ grid points over $5000$ generations, with error bars denoting standard deviations. The grey dashed line represents the threshold employed to compute the characteristic length scale $l$, defining the average size of the typical spatial domain occupied by each species. The simulation parameters are consistent with those illustrated in Fig.~\ref{fig2}, with $\gamma$ varying between $0$ and $1$ in increments of $\Delta=0.1$, as shown in the inset figure.

As $\gamma$ rises, highlighting the influence of antagonistic mortality, spatial correlation among individuals of the same species decreases. The inset reveals a roughly $10\%$ decrease in the characteristic length scale as $\gamma$ varies from $0$ to $1$. This phenomenon arises from a higher mortality rate, disrupting the cyclic variation inherent to the rock-paper-scissors rules and consequently enlarging the average size of spatial patterns occupied by each species.

\begin{figure}[t]
\centering
    \includegraphics[width=85mm]{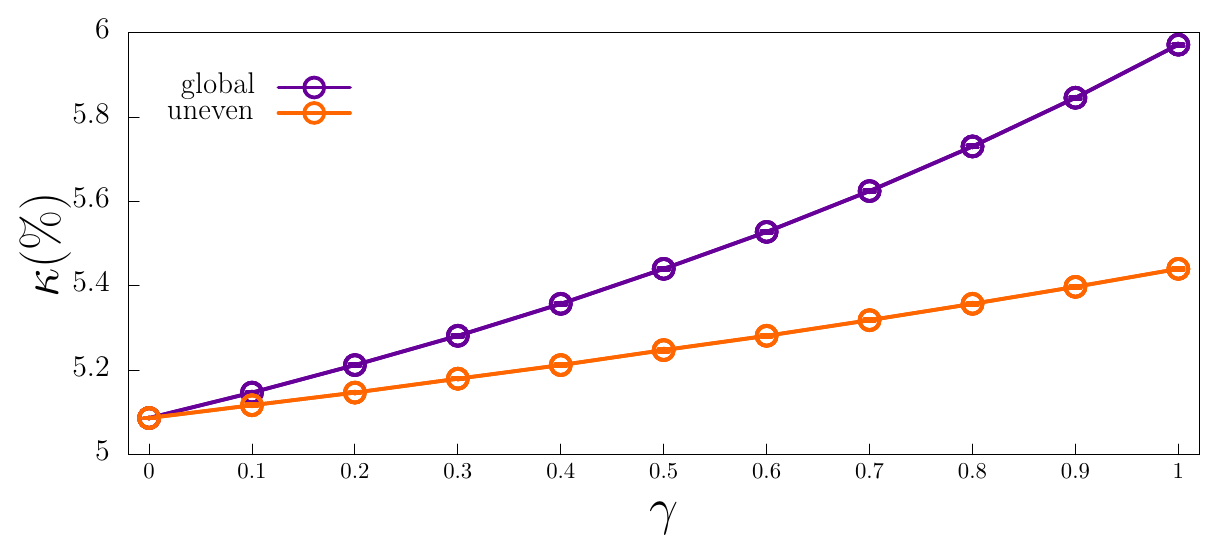}
\caption{Cure probability as a function of the antagonistic factor. The results were averaged over $100$ simulations running in lattices with $500^2$ grid sites until $t=5000$ generations. The error bars show the standard deviation. The orange line illustrates the scenario where the antagonistic coinfection impacts the mortality of both diseases. In contrast, the purple line represents the asymmetric case, where only the mortality of disease $1$ is influenced.}
	\label{fig5}
\end{figure}

\section{The impact of the antagonistic coinfection at individual and population levels}
\label{sec5}

Let us now examine the impact of antagonistic mortality on organisms' susceptibility to becoming infected and the probability of being cured in two scenarios:
\begin{enumerate}
\item
Global antagonistic coinfection: $0.0\,\leq \gamma_1 = \gamma_2 =\gamma \leq 1.0$, in intervals of $\Delta\,\gamma =0.1$; 
\item
Uneven antagonistic coinfection: $0.0\,\leq \gamma_1 \leq 1.0$, in intervals of $\Delta\,\gamma_1 =0.1$ and $\gamma_2=0.0$.
\end{enumerate}
The other parameters are the same as in the previous sections. The sets of simulations consists of $100$ runs starting from different initial conditions in lattices with $500^2$ grid lattices running until $5000$ generations. We exclude the first half of each implementation, thus avoiding the accentuated fluctuations inherent to the initial stage of pattern formation \cite{Moura}.

We first compute the average value of cure probability, $\kappa(t)$, defined as the probability of a sick organism being cured by one of the diseases per unit of time, respectively. 
This is computed by following the steps:
i) counting the total number of sick individuals of species $i$ that are infected with a specific disease  when each generation begins; ii) computing the number of ill individuals of species $i$ that are cured by the respective disease during the 
generation; 
iii) calculating the cure probability, defined as the ratio between the number of cured individuals and the initial number of sick individuals of species $i$.

Then, we calculate the mean infection risk, $\chi(t)$, defined as the probability of a healthy individual being infected by one of the pathogens per unit of time. To achieve this, the algorithm follows these steps:
i) counting the total number of healthy individuals at the start of each generation; ii) computing the number of healthy individuals of species $i$ that are contaminated (virus $1$ or virus $2$) during the generation; 
iii) calculating the infection risk, as the ratio of the number of infected individuals and the initial number of healthy individuals of species $i$.

Furthermore, to quantify the influence of the antagonistic coinfection on population dynamics, we compute the density of organisms of the same species $\rho$.  Owing to the symmetry of the rock-paper-scissors game, the results are identical for every species; thus, we utilise the data from species $1$ to compute the species density.

\begin{figure}
\centering
    \includegraphics[width=85mm]{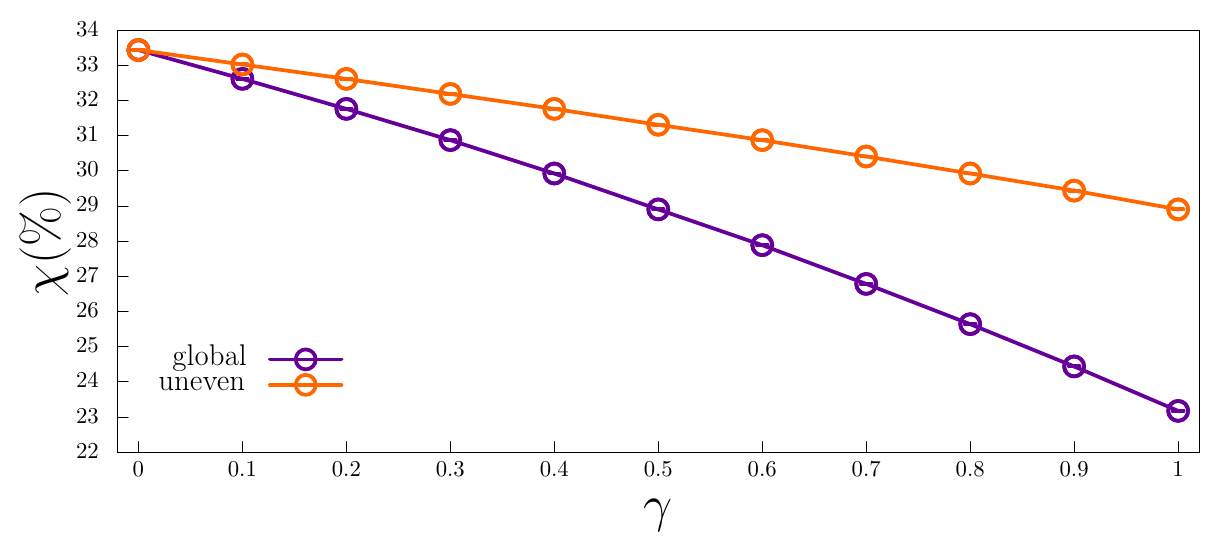}
\caption{Organisms' infection risk as a function of the coinfection antagonistic factor. The results were averaged from sets of $100$ simulations running in lattices with $500^2$ grid sites until $t=5000$ generations. The orange line depicts the scenario where the antagonistic coinfection affects the mortality of both diseases, while the purple line represents the uneven case where only the mortality of disease $1$ is influenced. The error bars indicate the standard deviation.}
	\label{fig6}
\end{figure}

\begin{figure}[t]
\centering
    \includegraphics[width=85mm]{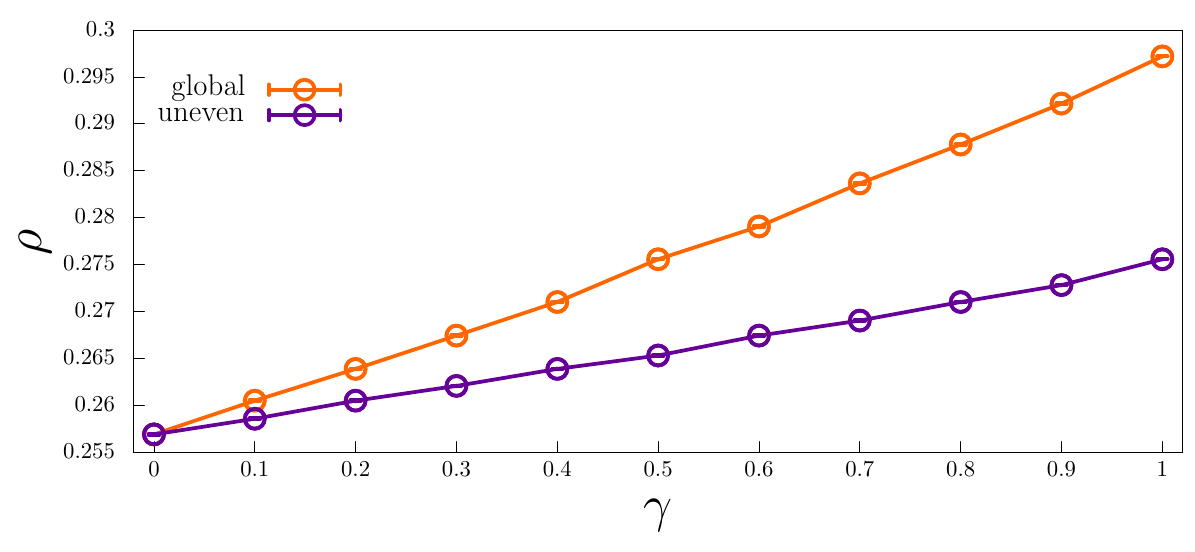}
\caption{Species density as a function of the coinfection antagonistic factor. The results were averaged from sets of $100$ simulations running in lattices with $500^2$ grid sites until $t=5000$ generations. The orange line depicts the scenario where the antagonistic coinfection interferes with the mortality of both diseases. In contrast, the purple line represents the uneven case, where only the mortality of disease $1$ is reduced. The error bars indicate the standard deviation.}
	\label{fig7}
\end{figure}

Figures \ref{fig5}, \ref{fig6}, and  \ref{fig7} show how the mean cure probability, the organisms' infection risk, and the species density vary with the antagonistic factor $\gamma$. The orange and purple lines represent the results for the global and uneven scenarios; the error bars show the standard deviation. 

According to our findings, as $\gamma$ grows and the coinfection becomes more antagonistic, the likelihood of coinfected hosts dying are reduced. This suggests that organisms gain additional time to mount immune responses against the pathogens responsible for the diseases, thus increasing the probability of being cured as $\gamma \to 1$. Furthermore, if the antagonistic coinfection is uneven, affecting only the development of disease $1$, coinfected organisms benefit by living longer and have more opportunities to recover from both diseases. This increases the cure probability for both diseases as $\gamma$ increases, regardless of whether the antagonistic coinfection is global or uneven.
Besides, our findings indicate that the maximum rise in cure probability is approximately $5.42\%$ for uneven antagonism and $5.95\%$ for global antagonism, as appear in Fig.~\ref{fig5} for $\gamma=1$. 
 This indicates that, relative to the baseline case with no antagonistic effect ($\gamma=0$), the maximum relative 
increase in the probability of being cured, which occurs for $\gamma=1$, is approximately $6.69\%$ if the antagonistic effect in the disease mortality is uneven while it reaches $17.52\%$ for the symmetric antagonism.

Figure \ref{fig6} shows that the organisms' infection risk decreases as antagonistic coinfection becomes more pronounced, with a reduction of approximately $26.27\%$ for total antagonism compared to the non-antagonistic coinfection case, compared to the non-antagonistic scenario. This phenomenon arises because, as coinfected organisms are less likely to die from one of the diseases, which increases the probability of cure. Therefore, new healthy individuals mostly appear when empty spaces are created through selection interactions, increasing the pool of susceptible organisms available for reinfection. 

Regarding species densities, as the chances of coinfected hosts dying decrease ($\gamma \to 1$), the benefits to the species population become more significant. The results show that the increase in species densities is approximately twice as large for global antagonism compared to uneven antagonism, with a maximum growth of approximately $11.32\%$ in the scenario of total antagonism of concurrent pathogen contamination, relative to the baseline case with no antagonistic effect ($\gamma=0$), as shown in Fig.~\ref{fig7}.


\section{Impact of the mobility restriction strategies in the organisms' life expectancy}
\label{sec6}

We now investigate how the implementation of mobility restrictions as a behavioural strategy can mitigate the impacts of deaths caused by disease infections, particularly in scenarios involving antagonistic coinfection. We expand the model presented in Ref. \cite{plasticity}, which was used to investigate the role of the mobility reduction during a single epidemic.  
In our stochastic simulation, each time an organism is randomly chosen to move, there exists a chance it might intentionally choose not to move, reflecting self-preservation tactics represented by $\nu$, where $\nu$ is the mobility restriction factor, a real parameter in the range $0\,\leq\,\nu\,\leq\,1$. This effectively reduces the probability of an organism moving within the system by $1-\nu$.

Our research aims to determine whether mobility restriction can positively influence organisms' life expectancy by reducing the likelihood of infection from lethal diseases. This is particularly relevant in the context of two concurrent epidemics, which interact within the system and thereby enhance the survival probability for individuals coinfected with both diseases compared to those infected with only one.

This study is systematically structured into three main components, each designed to enhance our understanding
of the impact of mobility restrictions on disease infections and survival rates in the context of antagonistic coinfection:
\begin{enumerate}
\item
Quantifying the variation in infection risk based on the mobility restriction factor.
\item
Evaluating the variation in the probability of organism mortality in the spatial game relative the mobility restriction factor.
\item
Analysing the dependence of organisms' expected lifetime on the mobility limitation factor.
\end{enumerate}
\subsection{The impact of mobility restriction in the organisms' infection risk}
To investigate the relationship between infection risk and the behavioural strategy of mobility restriction, we performed simulations using varying values of $\nu$ - we consider a global antagonism where both diseases are affected by reduced mortality.
We ran ensembles of 100 simulations for each $\nu$ value, ranging from $0$ to $1$ in intervals of $\Delta\nu=0.1$. Additionally, we introduced the relative variation of infection risk, denoted as $\tilde{\xi}$, calculated as $\tilde{\xi}=(\chi-\chi_0)/\chi_0$, where $\xi_0$ represents the infection risk in the absence of any mobility restriction strategy ($\nu=0$).

\begin{figure}
\centering
    \includegraphics[width=85mm]{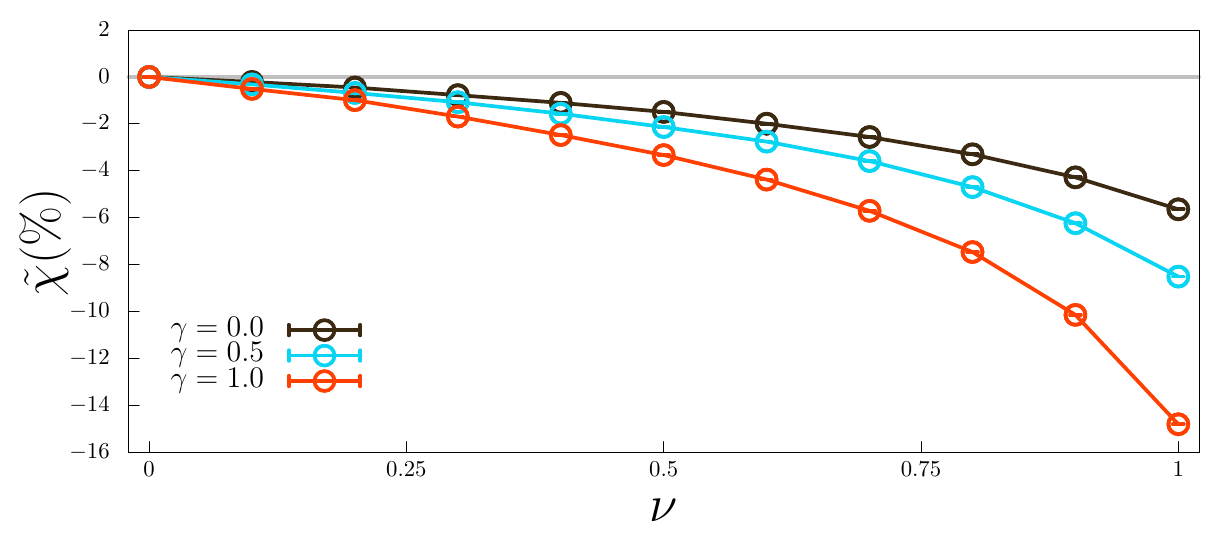}
\caption{Relative variation of the organisms' infection risk as a function of the mobility restriction factor for several levels of antagonistic coinfection. The results were averaged over $100$ simulations running in lattices with $500^2$ grid sites until $t=5000$ generations. The error bars show the standard deviation. Brown, cyan, and red lines depict the results for $\gamma=0.0$, $\gamma=0.5$, and $\gamma=1.0$, respectively. }
	\label{fig8}
\end{figure}

The results are depicted in Fig.~\ref{fig8}, where brown, cyan, and red lines correspond to scenarios with antagonistic mortality set at $\gamma=0.0$, $\gamma=0.5$, and $\gamma=1.0$, respectively. Error bars represent standard deviations.
The results suggest that reducing dispersal decreases the risk of healthy organisms becoming infected, irrespective of the mortality rate among sick individuals. Significantly, as mobility decreases, the infection risk diminishes because organisms
experience reduced exposure to potential viral vectors in their vicinity.

Additionally, we observe that the effectiveness of mobility restriction becomes more significant for antagonistic coinfection. According to Fig.~\ref{fig8}, complete restriction of dispersal reduces the infection risk by $5.5\%$ in non-antagonistic coinfection ($\gamma=0.0$). This reduction increases to $8.5\%$ when the probability of coinfected individuals dying from either disease is halved and reaches approximately $14.5\%$ in the case of total antagonism ($\gamma=1.0$).
\subsection{The influence of mobility limitation in the organisms' selection risk}
The next step is to examine whether the mobility restriction strategy, designed to reduce the risk of individuals contracting diseases, 
also influences the likelihood of individuals being eliminated through selection in the spatial rock-paper-scissors game. To achieve this, we calculate the selection risk, denoted as $\zeta(t)$, which represents the probability of an individual being eliminated within one generation of the spatial game, as introduced in Ref. \cite{Moura}:
i) at the beginning of each generation, we count the total number of individuals of species $i$ (both healthy and infected organisms).
ii) we determine the number of individuals of species $i$ eliminated by organisms of species $i-1$ during the generation;
iii) we compute the selection risk $\zeta_i$, where $i=1,2,3$, as the ratio between the number of eliminated individuals and the initial number of individuals of species $i$.
Finally, we compute the relative variation of selection risk, denoted as $\tilde{\zeta}$, calculated as $\tilde{\zeta}=(\zeta-\zeta_0)/\zeta_0$, where $\zeta_0$ represents the selection risk in the absence of any mobility restriction strategy ($\nu=0$).

\begin{figure}[t]
\centering
    \includegraphics[width=85mm]{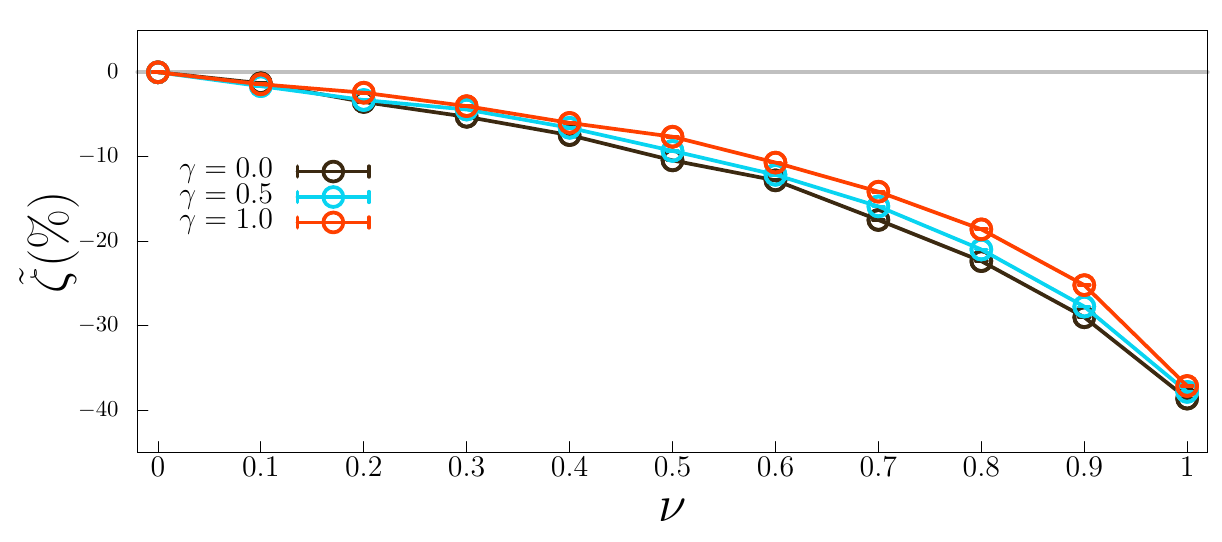}
\caption{Relative variation of the organisms' selection risk as a function of the mobility restriction factor for various antagonistic coinfection scenarios. The results were averaged from sets of $100$ simulations running in lattices with $500^2$ grid sites until $t=5000$ generations. The error bars show the standard deviation. Red, cyan, and brown lines depict the results for $\gamma=0.0$, $\gamma=0.5$, and $\gamma=1.0$, respectively. }
	\label{fig9}
\end{figure}

Subsequently, we run simulations with varying mobility restriction factor. For each $\nu$ value, we perform ensembles of 100 simulations, ranging from $0$ to $1$ in intervals of $\Delta\nu=0.1$. The outcomes are depicted in Fig.~\ref{fig9}, with brown, cyan, and red lines corresponding to scenarios with antagonistic mortality set at $\gamma=0.0$, $\gamma=0.5$, and $\gamma=1.0$, respectively. The error bars indicate standard deviations.

Our findings indicate that irrespective of mortality among infected individuals, reducing mobility decreases the likelihood of organisms being targeted for elimination. Notably, as individual mobility decreases, the reduction in selection risk becomes increasingly pronounced.
Furthermore, we observe that the effectiveness of mobility restriction becomes more pronounced 
as the level of antagonism in coinfected hosts' disease mortality decreases. For instance, at $\nu=0.8$, the reduction in selection risk reaches approximately $17.5\%$, $21.5\%$, and $22.5\%$ for $\gamma=0.0$, $\gamma=0.5$, and $\gamma=1.0$, respectively.

\subsection{Prolonging of organisms' expected survival time}
In our final analysis, we compute organisms' expected survival time, $\tau$, as a function of the mobility restriction factor. This involves evaluating the integral $\int_{0}^{\infty} S_{i} \, dt$, where the survival probability, $S_{i}(\nu)$, is determined by $S_{i}(\nu) = 1 - \omega(\nu)$, with $\omega$ representing the probability of individuals dying within one generation - either being eliminated  in the spatial game or succumbing to one of the lethal diseases spreading during concurrent epidemics.

\begin{figure}[t]
\centering
\includegraphics[width=85mm]{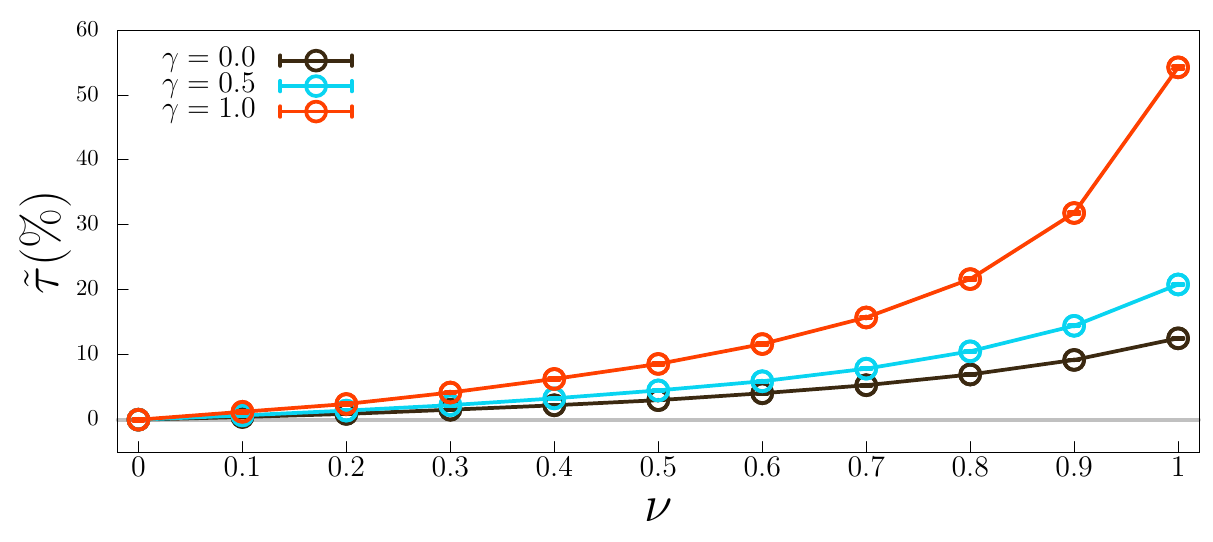}
\caption{Relative variation of the organisms' survival time as a function of the mobility restriction factor for various antagonistic coinfection scenarios. The results were averaged from sets of $100$ simulations running in lattices with $500^2$ grid sites until $t=5000$ generations. The error bars show the standard deviation. Red, cyan, and brown lines depict the results for $\gamma=0.0$, $\gamma=0.5$, and $\gamma=1.0$, respectively.}
\label{fig12}
\end{figure}

Fig.\ref{fig12} depicts the average survival time of organisms concerning the mobility restriction factor, based on $100$ simulations on lattices comprising $500^2$ grid sites over $5000$ generations. The error bars indicate standard deviations. The brown, cyan, and red lines correspond to results for antagonistic mortality factors $\gamma=0.0$, $\gamma=0.5$, and $\gamma=1.0$, respectively.
Subsequently, we compute the relative variation of organisms' expected survival time, denoted as $\tilde{\tau}$, calculated as $\tilde{\tau}=(\tau-\tau_0)/\tau_0$, where $\tau_0$ represents the expected survival time in the absence of any mobility restriction strategy ($\nu=0$).

Our analysis demonstrates that implementing mobility restrictions increases organisms' lifespans, with gains in life expectancy growing exponentially as less mobile organisms prevail. Additionally, we find that the effectiveness of the behavioural strategy of restricting mobility is enhanced when coinfection induces antagonistic effects on disease mortality. As shown in Fig.~\ref{fig12}, reducing individuals' dispersal may elevate organisms' chances of living longer by approximately $12.5\%$ in scenarios without antagonistic coinfection. This increase in life expectancy due to mobility restriction significantly rises to roughly $21.5\%$ for $\gamma=0.5$, reaching maximum benefits of approximately $54\%$ for $\gamma=1.0$.


\section{Discussion and Conclusions}
\label{sec7}
We investigate the interplay between antagonistic coinfection and mobility restriction strategies in a stochastic rock-paper-scissors model involving two concurrent epidemics. Our stochastic simulations cover all possible scenarios: the presence of one pathogen inhibiting the development of the disease caused by the other virus at varying levels, thereby reducing the likelihood of organisms dying from $0\%$ (no antagonistic mortality) to $100\%$ (total antagonism). We also examine the effects of uneven antagonistic mortality, where coinfection only slows the progression of one disease.

Our findings demonstrate that antagonistic coinfection significantly reduces mortality rates among coinfected individuals. This enhances the probability of sick individuals being cured by up to $17.52\%$ in cases of global antagonism and decreases the likelihood of healthy individuals becoming infected.
At the population level, antagonism benefits species populations, with growth reaching $11.32\%$ for global antagonism compared to $5.42\%$ for uneven antagonism. Studying the spatial patterns, we observe that antagonism disrupts cyclic dominance, producing fragmented spatial domains with a characteristic length scale that decreases as antagonistic effects grow.

Additionally, our results indicate that mobility restriction represents an effective behavioural strategy, further mitigating disease spread and enhancing survival. Limiting dispersal reduces infection risk, ranging from $5.5\%$ (non-antagonistic case) to $14.5\%$ (total antagonism). Mobility restriction also diminishes selection risks, improving survival through reduced exposure to pathogens and competitors. For example, restricting mobility by $80\%$ decreases selection risk by $17.5\%$–$22.5\%$, depending on the degree of antagonistic mortality. Therefore, the combination of reduced mobility and antagonistic coinfection amplifies the protective effects against concurrent epidemics, underscoring the dual role of mobility restriction in mitigating disease transmission and competitive elimination.

Expected survival time increases rapidly as mobility decreases, highlighting the effectiveness of mobility restriction in prolonging organisms' life expectancy may grow by $54\%$ under total antagonism. Furthermore, antagonistic coinfection amplifies these benefits by reducing mortality rates and promoting longer lifespans, particularly for coinfected hosts.

Our study shows the importance of integrating antagonistic coinfection and mobility restriction into ecological models. The synergy between mobility restriction and antagonistic interactions is pivotal in shaping individual-level outcomes and population dynamics. Our findings may be helpful to the design of environmental and public health interventions can combine behavioural adaptations and biological interactions to enhance resilience against infectious threats. Future research could extend our model to investigate long-term evolutionary outcomes and scenarios involving multiple interacting pathogens, incorporating adaptive behaviours or variable immunity levels.


\section*{Acknowledgments}
We thank CNPq/Brazil, ECT/UFRN, FAPERN/RN, IBED/UvA, ADSAI/Zuyd and Brightlands Smart Services Campus for financial and technical support.

\vspace{0.5cm}

\bibliography{COI}

\begin{thebibliography}{10}
\expandafter\ifx\csname url\endcsname\relax
  \def\url#1{\texttt{#1}}\fi
\expandafter\ifx\csname urlprefix\endcsname\relax\def\urlprefix{URL }\fi
\expandafter\ifx\csname href\endcsname\relax
  \def\href#1#2{#2} \def\path#1{#1}\fi

\bibitem{Coli}
B.~Kerr, M.~A. Riley, M.~W. Feldman, B.~J.~M. Bohannan, Local dispersal
  promotes biodiversity in a real-life game of rock–paper–scissors, Nature
  418 (2002) 171.

\bibitem{Allelopathy}
R.~Durret, S.~Levin, Allelopathy in spatially distributed populations, J.
  Theor. Biol. 185 (1997) 165--171.

\bibitem{bacteria}
B.~C. Kirkup, M.~A. Riley, Antibiotic-mediated antagonism leads to a bacterial
  game of rock-paper-scissors in vivo, Nature 428 (2004) 412--414.

\bibitem{lizards}
B.~Sinervo, C.~M. Lively, The rock-scissors-paper game and the evolution of
  alternative male strategies, Nature 380 (1996) 240--243.

\bibitem{coral}
J.~B.~C. Jackson, L.~Buss, The rock-scissors-paper game and the evolution of
  alternative male strategies, Proc. Natl Acad. Sci. USA 72 (1975) 5160--5163.

\bibitem{mobilia2}
T.~Reichenbach, M.~Mobilia, E.~Frey, Mobility promotes and jeopardizes
  biodiversity in rock-paper-scissors games, Nature 448 (2007) 1046--1049.

\bibitem{mobilia3}
T.~Reichenbach, M.~Mobilia, E.~Frey, Self-organization of mobile populations in
  cyclic competition, Journal of Theoretical Biology 254~(2) (2008) 368 -- 383.

\bibitem{uneven}
J.~Menezes, B.~Moura, T.~A. Pereira, Uneven rock-paper-scissors models:
  Patterns and coexistence, Europhysics Letters 126~(1) (2019) 18003.

\bibitem{tanimoto2}
K.~A. Kabir, J.~Tanimoto, The role of pairwise nonlinear evolutionary dynamics
  in the rock–paper–scissors game with noise, Applied Mathematics and
  Computation 394 (2021) 125767.

\bibitem{park2023}
J.~Park, B.~Jang, Role of adaptive intraspecific competition on collective
  behavior in the rock–paper–scissors game, Chaos, Solitons \& Fractals 171
  (2023) 113448.

\bibitem{rps-ambush}
R.~Barbalho, S.~Rodrigues, M.~Tenorio, J.~Menezes, Ambush strategy enhances
  organisms’ performance in rock–paper–scissors games, BioSystems 240
  (2024) 105229.

\bibitem{rps-scripta}
J.~Menezes, E.~Rangel, Reproduction-mobility trade-off in rock-paper-scissors
  models in changing environmental conditions, Physica Scripta 99~(4) (2024)
  045235.

\bibitem{social1}
E.~Du, E.~Chen, J.~Liu, C.~Zheng, How do social media and individual behaviors
  affect epidemic transmission and control?, Science of The Total Environment
  761 (2021) 144114.

\bibitem{disease4}
A.~M. Dunn, M.~E. Torchin, M.~J. Hatcher, P.~M. Kotanen, D.~M. Blumenthal,
  J.~E. Byers, C.~A. Coon, V.~M. Frankel, R.~D. Holt, R.~A. Hufbauer, A.~R.
  Kanarek, K.~A. Schierenbeck, L.~M. Wolfe, S.~E. Perkins, Indirect effects of
  parasites in invasions, Functional Ecology 26~(6) (2012) 1262--1274.

\bibitem{disease3}
M.~J. Young, N.~H. Fefferman, The dynamics of disease mediated invasions by
  hosts with immune reproductive tradeoff, Scientific Reports 12 (2022) 4108.

\bibitem{disease2}
T.~Nagatani, G.~Ichinose, K.~ichi Tainaka, Epidemics of random walkers in
  metapopulation model for complete, cycle, and star graphs, Journal of
  Theoretical Biology 450 (2018) 66--75.

\bibitem{socialdist}
T.~C. Reluga, Game theory of social distancing in response to an epidemic, PLoS
  Comput. Biol. 6~(5) (2010) e1000793.

\bibitem{soc}
S.~Stockmaier, N.~Stroeymeyt, S.~E. C., H.~D. M., L.~A. Meyers, D.~I. Bolnick,
  Infectious diseases and social distancing in nature, Science 371~(6533)
  (2021) eabc8881.

\bibitem{plasticity2}
N.~Stroeymeyt, A.~V. Grasse, A.~Crespi, D.~P. Mersch, S.~Cremer, L.~Keller,
  Social network plasticity decreases disease transmission in a eusocial
  insect, Science 362~(6417) (2018) 941--945.

\bibitem{mr0}
G.~Dimarco, G.~Toscani, M.~Zanella, Optimal control of epidemic spreading in
  the presence of social heterogeneity, Philosophical Transactions of the Royal
  Society A: Mathematical, Physical and Engineering Sciences 380~(2224) (2022)
  20210160.

\bibitem{mr1}
Q.~Shao, D.~Han, Epidemic spreading in metapopulation networks with
  heterogeneous mobility rates, Applied Mathematics and Computation 412 (2022)
  126559.

\bibitem{mr2}
P.~Edsberg~Møllgaard, S.~Lehmann, L.~Alessandretti, Understanding components
  of mobility during the covid-19 pandemic, Philosophical Transactions of the
  Royal Society A: Mathematical, Physical and Engineering Sciences 380~(2214)
  (2022) 20210118.

\bibitem{10.1371/journal.pone.0254403}
T.~Oka, W.~Wei, D.~Zhu, The effect of human mobility restrictions on the
  covid-19 transmission network in china, PLOS ONE 16~(7) (2021) 1--16.

\bibitem{CAPAROGLU2021111246}
Ömer Faruk~Çaparoğlu, Y.~Ok, M.~Tutam, To restrict or not to restrict? use
  of artificial neural network to evaluate the effectiveness of mitigation
  policies: A case study of turkey, Chaos, Solitons \& Fractals 151 (2021)
  111246.

\bibitem{Gene}
V.~Papanikolaou, A.~Chrysovergis, V.~Ragos, E.~Tsiambas, S.~Katsinis,
  A.~Manoli, S.~Papouliakos, D.~Roukas, S.~Mastronikolis, D.~Peschos,
  A.~Batistatou, E.~Kyrodimos, N.~Mastronikolis, From delta to omicron:
  S1-rbd/s2 mutation/deletion equilibrium in sars-cov-2 defined variants, Gene
  814 (2022) 146134.

\bibitem{mutating1}
M.~Becerra-Flores, T.~Cardozo, Sars-cov-2 viral spike g614 mutation exhibits
  higher case fatality rate, International Journal of Clinical Practice 74~(8)
  (2020) e13525.

\bibitem{plasticity1}
A.~Bridier, P.~C. Piard, J-C.~and, S.~Labarthe, F.~Dubois-Brissonnet,
  R.~Briandet, Spatial organization plasticity as an adaptive driver of surface
  microbial communities, Front. Microbiol 8 (2017) 1364.

\bibitem{synergy}
J.~Menezes, E.~Rangel, Spatial dynamics of synergistic coinfection in
  rock-paper-scissors models, Chaos: An Interdisciplinary Journal of Nonlinear
  Science 33~(9) (2023) 093115.

\bibitem{coifectedhiv}
L.~J. Abu-Raddad, P.~Patnaik, J.~G. Kublin, Dual infection with hiv and malaria
  fuels the spread of both diseases in sub-saharan africa, Science 314~(5805)
  (2006) 1603--1606.

\bibitem{antag2}
A.~Buckling, P.~B. Rainey, Antagonistic coevolution between a bacterium and a
  bacteriophage, Proceedings of the Royal Society of London. Series B:
  Biological Sciences 269~(1494) (2002) 931--936.

\bibitem{antag3}
M.~N.A., The influence of parasite infections on host immunity to co-infection
  with other pathogens, Front. Immunol. 9 (2018) 2579.

\bibitem{antag4}
S.~Manna, J.~McAuley, J.~Jacobson, C.~D. Nguyen, M.~A. Ullah, I.~Sebina,
  V.~Williamson, E.~K. Mulholland, O.~Wijburg, S.~Phipps, C.~Satzke, Synergism
  and antagonism of bacterial-viral coinfection in the upper respiratory tract,
  mSphere 7~(1) (2022) e00984--21.

\bibitem{antag1}
S.-S. Shen, X.-Y. Qu, W.-Z. Zhang, J.~Li, Z.-Y. Lv, Infection against
  infection: parasite antagonism against parasites, viruses and bacteria,
  Infectious Diseases of Poverty 49~(8) (2019) 126559.

\bibitem{epidemicbook}
F.~M. Snowden, Epidemics and society: from the black death to the present, Yale
  University Press, New Haven and London, 2019.

\bibitem{tanimoto}
J.~Tanimoto, Sociophysics Approach to Epidemics, Springer,, Singapore, 2021.

\bibitem{leonard}
R.~M. May, W.~J. Leonard, Nonlinear aspects of competition between three
  species, SIAM J. Appl. Math. 29 (1975) 243--253.

\bibitem{plasticity}
J.~Menezes, S.~Batista, E.~Rangel, Spatial organisation plasticity reduces
  disease infection risk in rock–paper–scissors models, Biosystems 221
  (2022) 104777.

\bibitem{mobilityrestrictions}
J.~Menezes, Mobility restrictions in response to local epidemic outbreaks in
  rock-paper-scissors models, Journal of Physics: Complexity 5~(1) (2024)
  015018.

\bibitem{Moura}
B.~Moura, J.~Menezes, Behavioural movement strategies in cyclic models,
  Scientific Reports 11 (2021) 6413.

\end{thebibliography}

\end{document}